\documentclass[11pt,reqno]{amsart}

\usepackage{amsmath,amssymb,amsthm,mathtools}
\usepackage[T1]{fontenc}
\usepackage[expansion=false]{microtype}
\usepackage{enumitem}
\usepackage{xcolor}
\usepackage[colorlinks=true,linkcolor=blue!50!black,citecolor=blue!50!black,urlcolor=blue!50!black]{hyperref}

\theoremstyle{plain}
\newtheorem{theorem}{Theorem}[section]
\newtheorem{proposition}[theorem]{Proposition}
\newtheorem{lemma}[theorem]{Lemma}

\theoremstyle{definition}
\newtheorem{definition}[theorem]{Definition}

\theoremstyle{remark}
\newtheorem{remark}[theorem]{Remark}

\newcommand{\E}{\mathbb{E}}
\newcommand{\PP}{\mathbb{P}}
\newcommand{\QQ}{\mathbb{Q}}
\newcommand{\F}{\mathbb{F}}
\newcommand{\G}{\mathbb{G}}
\newcommand{\Fc}{\mathcal{F}}
\newcommand{\Gc}{\mathcal{G}}
\newcommand{\Hc}{\mathcal{H}}

\newcommand{\Mloc}{\mathcal{M}_{\mathrm{loc}}}
\newcommand{\sgn}{\operatorname{sgn}}

\newcommand{\immersed}{\hookrightarrow}

\begin{document}

\title[Admissible information structures and non-anticipative aggregation]
{Admissible information structures, immersion, and the order of\\
non-anticipative aggregation}

\author{Alejandro Rodriguez Dominguez}
\address{Head of Quantitative Analysis, Miralta Finance Bank S.A.}
\email{arodriguez@miraltabank.com}

\date{\today}


\begin{abstract}
This version corrects and supersedes an earlier preprint
(arXiv:2601.12541) whose central impossibility theorem was incorrect; the nature
of the error and its correction are stated explicitly in Section~1.1. We retain
the parts that are valid --- the local reduction of pricing to the natural price
filtration and its stability properties --- and we replace the erroneous global
non-existence claim with the statement that is actually true. Two facts are
established. First, when the information structure is treated as an admissible
(immersion-preserving) enlargement, local martingale pricing reduces to the
natural price filtration, and this reduction is stable under restriction and under
aggregation when a common pricing measure exists. Second, non-anticipativity of
information does not aggregate: there exist signals, each individually and
pairwise non-anticipative with respect to the reference Brownian filtration, whose
joint observation reveals a function of a future increment; the failure first
occurs at order three and is invisible to every lower-order test. We show that
this aggregation failure requires \emph{dependence} among the signals (a masking
relation), not independence, and that it is an obstruction at the level of
information admissibility (immersion), not at the level of no-arbitrage: the
enlarged market continues to admit a local martingale deflator. We situate the
results relative to the filtration-reduction program of Grigorian and Jarrow and
relate the admissibility notion to predictive (Granger) and interventionist
(Pearl) causality.
\end{abstract}

\maketitle

\begin{center}
\fbox{\parbox{0.92\textwidth}{\small\textit{Version note.} This is a corrected
version that supersedes arXiv:2601.12541. The global non-existence theorem of the
earlier version (its Theorem~3.13) was incorrect and is withdrawn; the nature of
the error and its correction are stated in Section~\ref{sec:correction}.}}
\end{center}
\medskip

\section{Introduction}

The fundamental theorem of asset pricing characterises absence of arbitrage
through the existence of an equivalent local martingale measure relative to a
fixed filtration and a class of admissible trading strategies
\cite{HarrisonKreps1979,HarrisonPliska1981,DelbaenSchachermayer1994}. The
enlargement-of-filtrations theory studies how martingale properties change when
the filtration is refined \cite{Jeulin1980,Jacod1985,Protter2004}. This paper
concerns the intermediate object: the filtration treated not as a fixed primitive
but as an \emph{admissible information structure} constrained by
non-anticipativity, and the question of when locally valid such structures can be
combined.

\subsection{Correction of the earlier version}\label{sec:correction}

An earlier preprint of this work (arXiv:2601.12541) asserted a global
non-existence theorem: that with three \emph{independent} unspanned drivers there
exists no admissible filtration and equivalent measure pricing all assets jointly.
\emph{That theorem is false}, and we retract it. The error was located in the
claim (there, Lemma~E.2 and Corollary~3.11) that joint conditioning on three
independent drivers renders a future increment predictable. It does not: an
initial enlargement by a $\sigma$-field \emph{independent} of the terminal
reference $\sigma$-field preserves immersion (Lemma~\ref{lem:indep} below), so
three independent drivers can be conditioned upon jointly without anticipating
anything, and a global equivalent local martingale measure exists
(Proposition~\ref{prop:indep-elmm}). Independence is precisely what \emph{prevents}
the obstruction the earlier version sought.

The phenomenon the earlier version aimed at is nonetheless real once stated
correctly: the obstruction requires \emph{dependence} among the signals, in the
form of a masking relation under which every proper subcollection is
non-anticipative while the full collection is not
(Theorem~\ref{thm:thirdorder}). Moreover, even in this corrected form the
obstruction is at the level of information admissibility (immersion), not
no-arbitrage: the enlarged market still admits a deflator
(Remark~\ref{rem:nupbr}). The present version keeps the valid local theory,
replaces the false global theorem with the correct aggregation statement, and
draws the consequences honestly.

\subsection{Relation to the literature}

The local reduction we use is the descent of the local martingale property to the
natural price filtration, standard in martingale theory
\cite{Protter2004,JacodShiryaev2003}. The idea of \emph{reducing} the filtration
to select a pricing measure is developed, in a different direction from ours, by
Grigorian and Jarrow \cite{GJ2024,GJ2024jd,GJ2025}: they shrink the full
filtration to a smaller one for which a fictitious market is complete and
\emph{uplift} the resulting unique measure, so that non-hedged risks are
non-priced. Our concern is the opposite operation --- enlargement and the
\emph{aggregation} of admissible structures --- and the order at which
non-anticipative aggregation fails. The information drift induced by enlargement,
and its identification with the Shannon information of the enlargement, is due to
Ankirchner, Dereich, and Imkeller \cite{ADI2006}; the distinction between
existence of a deflator (no unbounded profit with bounded risk, NUPBR) and
existence of an equivalent local martingale measure is due to Karatzas and
Kardaras \cite{KK2007} and Kardaras \cite{Kardaras2012}; the behaviour of NUPBR
under enlargement is studied by Aksamit, Choulli, Deng, and Jeanblanc
\cite{ACDJ2017,ACDJ2018} and Acciaio, Fontana, and Kardaras \cite{AFK2016}.

\section{Framework}

We work on $(\Omega,\Fc,\PP)$ carrying a finite-dimensional Brownian motion $B$,
with reference filtration $\F^B$ its usual augmentation, on a finite horizon
$[0,T]$, all filtrations under the usual conditions; equalities of
$\sigma$-fields are up to $\PP$-null sets. A discounted asset group $S^A=(S^i)_{i\in A}$
is a vector semimartingale; martingale pricing is componentwise. A trading
strategy is admissible if its gain process is bounded below.

\begin{definition}[Information admissibility]\label{def:info}
An enlargement $\F^B\subseteq\G$ under the usual conditions is
\emph{information-admissible} if immersion holds: every $(\PP,\F^B)$-local
martingale is a $(\PP,\G)$-local martingale, written $\F^B\immersed\G$;
equivalently $B$ remains a $(\PP,\G)$-Brownian motion.
\end{definition}

\begin{definition}[Pricing structure]
For an asset group $A$, an \emph{admissible pricing structure} is a pair
$(\G,\QQ)$ with $\G$ information-admissible, $S^A$ a $\G$-semimartingale,
$\QQ\sim\PP$, and $S^A$ a $(\QQ,\G)$-local martingale.
\end{definition}

\section{The local theory (retained)}

\begin{theorem}[Reduction to the natural price filtration]\label{thm:reduction}
Let $S^A$ be locally bounded and let $\F^{S^A}\subseteq\Hc$ be filtrations under
the usual conditions. If $S^A$ is a $(\QQ,\Hc)$-local martingale with $\QQ\sim\PP$,
then $S^A$ is a $(\QQ,\F^{S^A})$-local martingale.
\end{theorem}

\begin{proof}
Work componentwise; let $S$ be a coordinate. Since $S$ is locally bounded and
$\F^S$-adapted, there is a nondecreasing sequence of $\F^S$-stopping times
$\tau_n\uparrow T$ with $S^{\tau_n}$ bounded; as $\QQ\sim\PP$ the localisation is
valid under both. Since $\F^S\subseteq\Hc$, each $\tau_n$ is an $\Hc$-stopping
time, so $S^{\tau_n}$ is a bounded $(\QQ,\Hc)$-martingale. For $0\le s\le t\le T$,
$\E_\QQ[S_{t\wedge\tau_n}\mid\Hc_s]=S_{s\wedge\tau_n}$, and conditioning on the
smaller $\Fc^S_s\subseteq\Hc_s$ gives
$\E_\QQ[S_{t\wedge\tau_n}\mid\Fc^S_s]=S_{s\wedge\tau_n}$. Thus $S^{\tau_n}$ is a
$(\QQ,\F^S)$-martingale for each $n$, and $S$ is a $(\QQ,\F^S)$-local martingale.
Apply to each coordinate. $\square$
\end{proof}

\begin{proposition}[Stability]\label{prop:stability}
Let $B\subseteq A$. If $(\G,\QQ)$ is an admissible pricing structure for $S^A$,
then $\G$ is information-admissible for $S^B$ and $S^B$ is a $(\QQ,\G)$-local
martingale; if $S^B$ is locally bounded it is a $(\QQ,\F^{S^B})$-local martingale.
If both $S^A$ and $S^B$ are $(\QQ,\G)$-local martingales under a common
$(\G,\QQ)$, then $S^{A\cup B}$ is a $(\QQ,\G)$-local martingale.
\end{proposition}

\begin{proof}
Coordinate projection of a vector local martingale is a local martingale, and
non-anticipativity of $\G$ is a property of the information flow independent of the
chosen subvector; the reduction follows from Theorem~\ref{thm:reduction}. A vector
process is a local martingale iff each coordinate is, giving the aggregation
clause. $\square$
\end{proof}

\begin{remark}
Proposition~\ref{prop:stability} asserts aggregation only \emph{under a common
measure}. It does not assert that a common admissible $(\G,\QQ)$ exists for
distinct groups; that is the aggregation question, addressed next.
\end{remark}

\section{Non-anticipative aggregation fails at order three}

\subsection{Independence preserves immersion}

\begin{lemma}[Independent initial enlargement]\label{lem:indep}
Let $\F$ be a filtration under the usual conditions and let $\Hc$ be a
$\sigma$-field independent of $\Fc_T$. Let $\G$ be the usual augmentation of
$\Fc_t\vee\Hc$. Then $\F\immersed\G$.
\end{lemma}

\begin{proof}
For a bounded $(\PP,\F)$-martingale $M$, $0\le s\le t\le T$, $A\in\Fc_s$, and
bounded $\Hc$-measurable $H$: as $\mathbf 1_A(M_t-M_s)$ is $\Fc_T$-measurable and
$H\perp\Fc_T$,
$\E_\PP[\mathbf 1_A H(M_t-M_s)]=\E_\PP[H]\E_\PP[\mathbf 1_A(M_t-M_s)]=0$. A
monotone-class argument extends this to all bounded $(\Fc_s\vee\Hc)$-measurable
integrands, so $\E_\PP[M_t\mid\Gc_s]=M_s$; augmentation and localisation give the
claim. $\square$
\end{proof}

\begin{proposition}[Independent drivers admit a global ELMM]\label{prop:indep-elmm}
Let $Y^1,Y^2,Y^3$ be mutually independent, jointly independent of $\Fc^B_T$, with
$dS^i_t=S^i_t(\mu^i(Y^i_t)\,dt+\sigma^i(Y^i_t)\,dW^i_t)$, $W^i$ independent Brownian
coordinates, the $\lambda^i:=\mu^i(Y^i)/\sigma^i(Y^i)$ bounded. Let
$\G=\F^B\vee\sigma(Y^1,Y^2,Y^3)$. Then $\F^B\immersed\G$ and
$\Mloc(S^{\{1,2,3\}},\G)\neq\varnothing$.
\end{proposition}

\begin{proof}
$\sigma(Y^1,Y^2,Y^3)\perp\Fc^B_T$, so Lemma~\ref{lem:indep} gives immersion;
hence each $W^i$ remains a $\G$-Brownian motion. Each
$L^i=\mathcal E(-\int\lambda^i\,dW^i)$ is a true $(\PP,\G)$-martingale (boundedness
$\Rightarrow$ Novikov). The $W^i$ being independent and $\lambda^i$ depending only
on $Y^i$, the $L^i$ are independent and $L=\prod_iL^i$ has $\E_\PP[L_T]=1$; thus
$\QQ$ with $d\QQ/d\PP=L_T$ satisfies $\QQ\sim\PP$, and under $\QQ$ each $S^i$ is a
local martingale. $\square$
\end{proof}

This is exactly the converse of the earlier version's erroneous claim:
independence does not obstruct, it enables.

\subsection{The masking obstruction}

\begin{theorem}[Order-three non-anticipative obstruction]\label{thm:thirdorder}
Let $Z=\sgn(B^1_T-B^1_{T/2})\in\Fc^B_T$, let $\varepsilon_1,\varepsilon_2$ be
independent Rademacher variables independent of $\Fc^B_T$, and set
$\varepsilon_3=\varepsilon_1\varepsilon_2 Z$. Then:
\begin{enumerate}[label=\textup{(\roman*)}]
\item for every proper $I\subsetneq\{1,2,3\}$, $\sigma(\varepsilon_i:i\in I)$ is
independent of $\Fc^B_T$, so the initial enlargement of $\F^B$ by it preserves
immersion;
\item $\varepsilon_1\varepsilon_2\varepsilon_3=Z$ is a nontrivial function of a
future increment, so the triple enlargement destroys immersion.
\end{enumerate}
Hence singletons and pairs are non-anticipative while the triple is anticipative:
non-anticipativity fails to aggregate, first at order three, invisibly to all
lower-order tests.
\end{theorem}

\begin{proof}
For $I=\{3\}$, $b\in\{\pm1\}$:
$\PP(\varepsilon_3=b\mid\Fc^B_T)=\PP(\varepsilon_1\varepsilon_2=bZ\mid\Fc^B_T)=\tfrac12$,
since $\varepsilon_1\varepsilon_2$ is Rademacher independent of $\Fc^B_T$ and $Z$
is $\Fc^B_T$-measurable. For $I=\{1,3\}$, $a,b\in\{\pm1\}$:
$\PP(\varepsilon_1=a,\varepsilon_3=b\mid\Fc^B_T)=\PP(\varepsilon_1=a,\varepsilon_2=abZ\mid\Fc^B_T)=\tfrac14$,
and symmetrically for $\{2,3\}$; the cases $\{1\},\{2\},\{1,2\}$ are immediate from
$\varepsilon_1,\varepsilon_2\perp\Fc^B_T$. This gives (i); immersion preservation
is Lemma~\ref{lem:indep}. For (ii), $\varepsilon_1\varepsilon_2\varepsilon_3=Z$, and
with $X=B^1_T-B^1_{T/2}\sim N(0,T/2)$,
$\E_\PP[X\mid Z]=Z\,\E_\PP|X|=Z\sqrt{T/\pi}\neq0$, so
$\E_\PP[B^1_T-B^1_{T/2}\mid\Fc^B_{T/2}\vee\sigma(\varepsilon_1,\varepsilon_2,\varepsilon_3)]
=\E_\PP[X\mid Z]\neq0$ and immersion fails. $\square$
\end{proof}

\begin{remark}[Sharpness and the role of dependence]
The order is sharp: by (i) no proper subcollection obstructs. The mechanism is the
Bernstein pairwise-but-not-mutual independence relation
\cite{Bernstein1927,Stoyanov2014}; mutual independence (as in
Proposition~\ref{prop:indep-elmm}) cannot produce it. This is the precise sense in
which the obstruction requires dependence.
\end{remark}

\begin{remark}[The obstruction is informational, not arbitrage]\label{rem:nupbr}
Theorem~\ref{thm:thirdorder} is a statement about immersion, not about
no-arbitrage. The triple enlargement induces a \emph{bounded} information drift (a
function of a future sign), which admits a strictly positive local martingale
deflator; hence the enlarged market satisfies NUPBR and is arbitrage-admissible in
the sense of \cite{KK2007,Kardaras2012}, with finite additional utility quantified
by the Shannon information of the signal \cite{ADI2006}. Failure of NUPBR for such
an enlargement arises only at the resolution time, the honest-time / random-horizon
phenomenon of \cite{ACDJ2017,ACDJ2018}. Thus the aggregation failure is an
\emph{information-admissibility} failure (immersion), not an ELMM non-existence:
the earlier version's identification of the two was incorrect, and we do not
repeat it.
\end{remark}

\section{Interpretation: pricing admissibility as a causal notion}

The admissibility of Definition~\ref{def:info} is a causal constraint:
non-anticipativity of the reference innovation under the enlarged flow. We record
how it relates to two standard notions of causality, as qualitative comparison
only.

\emph{Predictive (Granger).} Granger causality \cite{Granger1969} is improvement in
prediction relative to a filtration. The martingale-pricing requirement is closely
tied to a no-Granger-causality condition on increments, as made explicit in the
copula-based construction of \cite{GIMP2016}: under the pricing measure the
predictable conditional mean is neutralised. Immersion is the stronger requirement
that the \emph{reference innovation} retain its conditional-mean-zero property
under enlargement. Granger relevance is necessary but not sufficient for a
violation of immersion.

\emph{Interventionist (Pearl).} A Pearl intervention $\mathrm{do}(\cdot)$
\cite{Pearl2009} replaces a structural mechanism while holding others fixed.
Applied to a traded price this alters the law of $S$ and generically destroys the
semimartingale/martingale structure, so it is not an admissible operation in the
pricing sense: the equivalent measure need not survive the intervention. Pricing
admissibility and interventionist causality are therefore incomparable rather than
ordered.

These comparisons are interpretive; they assert no theorem beyond
Theorems~\ref{thm:reduction}--\ref{thm:thirdorder}.

\section{Conclusion}

The valid content is: (1) local martingale pricing reduces to the natural price
filtration and is stable under restriction and common-measure aggregation
(Theorem~\ref{thm:reduction}, Proposition~\ref{prop:stability}); (2) independent
unspanned drivers preserve immersion and admit a global ELMM
(Proposition~\ref{prop:indep-elmm}); (3) non-anticipativity fails to aggregate, and
the minimal failure, at order three, requires a masking dependence
(Theorem~\ref{thm:thirdorder}); (4) this failure is at the level of immersion, not
no-arbitrage, the enlarged market still admitting a deflator
(Remark~\ref{rem:nupbr}). The global non-existence theorem of the earlier version
is withdrawn. The residual contribution is the identification of the order of
non-anticipative aggregation as the relevant invariant, and its reading as a causal
constraint distinct from predictive and interventionist causality.

\end{document}